\documentclass[aps,showpacs,twocolumn,10pt,prl]{revtex4-1}

\usepackage{amsmath}
\usepackage{amsthm}
\usepackage{amsfonts}
\usepackage{amssymb}
\usepackage{graphicx}

\begin{document}
\title{Recollision scenario without tunneling : Role of the ionic core potential}

\author{A. Kamor$^{1,2}$, C. Chandre$^{2}$, T. Uzer$^{1}$, F. Mauger$^3$}
\affiliation{
$^{1}$ School of Physics, Georgia Institute of Technology, Atlanta, GA 30332-0430, USA \\
$^{2}$ Centre de Physique Th\'eorique, CNRS -- Aix-Marseille Universit\'e, Campus de Luminy, case 907, 13009 Marseille, France\\
$^3$ Laboratoire de Chimie Th\'eorique, Facult\'e des Sciences, Universit\'e de Sherbrooke, Sherbrooke, Qu\'ebec, Canada J1K 2R1
}


\begin{abstract}
The standard model of strong laser physics, the recollision scenario, omits the ionic core potential after tunneling. Strikingly, although the Coulomb interaction drives all stages of recollision, the maximum energy the electrons bring back to the core is found by ignoring it. We resolve this long-standing paradox by showing that this good agreement stems from a fortuitous cancellation at high intensities. Instead of the three step model, we find that the Coulomb interaction can be fully integrated into a purely classical scenario that explains recollisions without invoking tunneling.
\end{abstract}
\pacs{32.80.Rm, 05.45.Ac, 32.80.Fb} 
\maketitle

The advent of powerful and short laser pulses heralded a new era in atomic and molecular physics about three decades ago~\cite{Cork07,Krau09}. Irradiating targets such as atoms, molecules, or biological complexes with such intense laser pulses has become the tool of choice for resolving the structure of matter at unprecedented spatial and temporal scales~\cite{Itat04}. About twenty years ago, the theoretical framework for such processes was established and remains the state-of-the-art in strong field physics~\cite{Cork93,Scha93}. It centers on the recollision model which follows a ``three-step'' scenario: Electrons are first detached (presumably by tunneling) and absorb energy while following the laser only to be hurled back at the ionic core when the laser reverses direction where they can ionize more electrons~\cite{Beck12} or generate very high harmonics of the driving laser by high harmonic generation (HHG)~\cite{Lewe94}.  One decisive success of the recollision model is the theoretical prediction of the highest harmonic $N_{\rm max}$ generated during this recombination process (the so-called ``high harmonic cut-off''~\cite{Lhui93b}). For a linearly polarized laser, where the electric field is $E(t)=E_0 \sin(\omega t+\phi)$, the cut-off is given by the analytical expression~\cite{Cork93,Krau92}
$$
 \hbar N_{\rm max}\omega = \kappa_0 U_p +I_p,
$$      
where $ \kappa_0  U_p $ is the maximum energy brought back by the recolliding electron in terms of the ponderomotive energy $U_p=E_0^2/(4\omega^2)$, $I_p$ is the ionization potential, and  $\kappa_0\approx 3.17$ is given by $\kappa_0=2(1-\cos \phi_0)$ where the laser phase $\phi_0\approx 4.09$ is the first positive root of $ \cos \phi +2(1-\cos \phi)/\phi^2-2\sin\phi /\phi =0$~\cite{Lewe94}. Remarkably, although the Coulomb interaction drives all stages of recollision, the maximum return energy of the recolliding electrons is found by simply ignoring it. It turns out that electrons can extract more (or less) energy from the laser field depending on when they are launched and when they are detected~\cite{Scha93,Prot96,Sand99}, but the validity of $3.17~{ U_p}$ for electrons which start and end at the core  has been confirmed in HHG experiments on such a regular basis~\cite{Lhui93b,Lhui93a} in the last two decades that it is often forgotten that it is based on a radical assumption, namely neglecting the parent ion's Coulomb interaction after it creates a potential barrier through which the photoelectrons tunnel ~\cite{Pfei12,Shaf12b}. With this simplifying assumption, called the Strong Field Approximation (SFA)~\cite{Ivan05}, Newton's equations give a maximum return energy $ \kappa_0  U_p $. This is one of many situations in intense laser physics in which classical mechanics acts as a reliable guide to quantum simulations and experimental observations~\cite{Beck12}.

In reality, for typical HHG experiments, the Coulomb interaction cannot be neglected, even after tunneling, since it drives all kinds of intense-laser phenomena: It ``focuses'' spreading electronic wave packets~\cite{Barb96,Shaf13,Li13}, and it guides electrons towards the core in certain directions~\cite{Shaf12a}, to cite a few examples. So why can the maximum energy brought back to the core be found as if the Coulomb interaction did not exist? Here we resolve this paradox with a purely classical recollision scenario which fully incorporates the Coulomb interaction at every stage of the recollision process. It shows that at high intensities (above $10^{15}\ {\rm W}\cdot{\rm cm}^{-2}$) the SFA procedure gives the right HHG cut-off thanks to a cancellation but that at lower intensities the three-step scenario with the SFA is no longer valid.  
\begin{figure}
 \centering
 \includegraphics[width = \linewidth, type=png, read=.png, ext=.png]{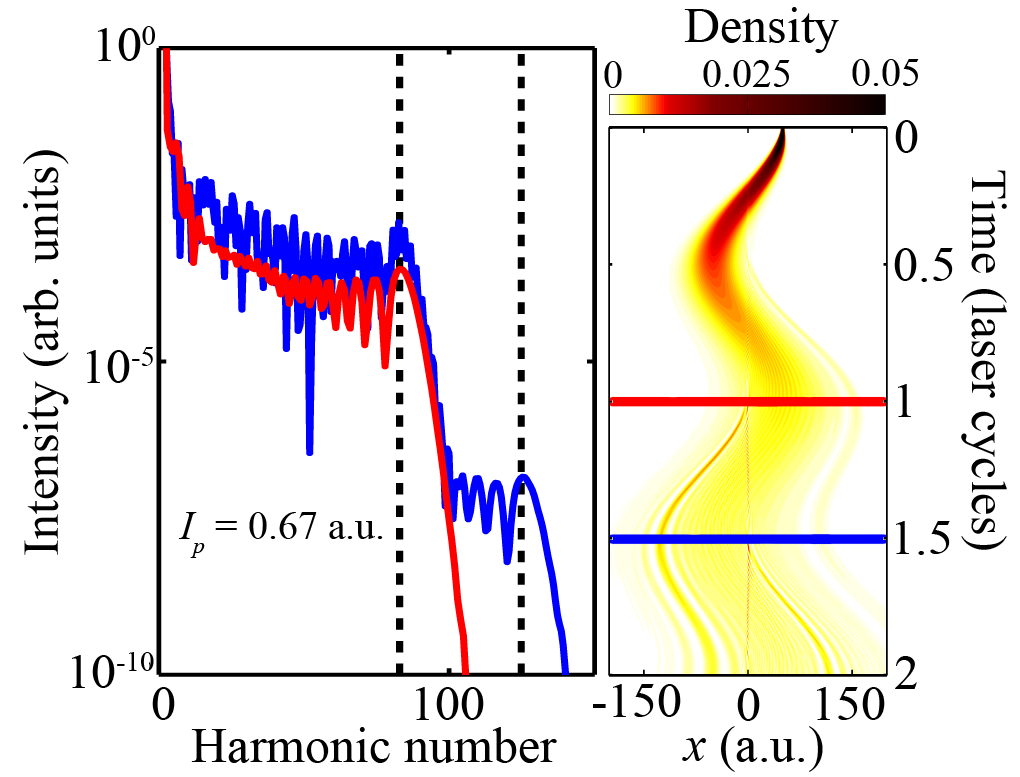}		 
\caption{ \label{fig:HHG} Left side: HHG spectra for the one-dimensional soft Coulomb potential at $I=10^{15} \ {\rm W}\cdot{\rm cm}^{-2}$ and a wavelength of 780 nm.  A generic atom is modelled with an ionization potential of $I_p = 0.67$ a.u.  The dashed lines correspond to 2 $U_p + I_p$ (left) and 3.17 $U_p + I_p$ (right).  The wave packet is launched at the quiver radius with the laser phased so that the wave-packet immediately returns to the core region.  The red curve is the HHG spectrum after one laser cycle while the blue curve is after one and a half laser cycles.  Right side: Density (color scale) of the wave packet as as a function of time (increasing in the downward direction) and position of the electron.  The red and blue lines show the wave packet at the times corresponding to the same colored curves on the left.}
\end{figure}
The prototypical figure we will interpret using our proposed recollision scenario is Fig.~\ref{fig:HHG} which depicts HHG spectra in a scattering simulation where the wave packet is initially launched at the quiver radius ($E_0 / \omega^2$) and the initial laser phase is chosen so that the wave packet returns to the ionic core in the next quarter laser cycle (see right panel of Fig.~\ref{fig:HHG}). The spectra have been computed by solving the one-dimensional Schr\"odinger equation with a soft-Coulomb potential~\cite{Java88} and a linearly polarized laser pulse in the near infrared regime. We observe one plateau after one laser cycle (red curve) and two plateaus after one and a half laser cycles (blue curve): The first plateau has a cut-off at $2~U_p+I_p$~\cite{Prot96,Sand99} and the second one at $3.17~ U_p+I_p$. In a different context, using Rydberg excited states, these two plateaus have been found as universal~\cite{Bled13}. Using our purely classical recollision scenario, we provide an explanation for these two plateaus and their universality in what follows. 

\begin{figure}
 \centering
 \includegraphics[width = \linewidth, type=png, read=.png, ext=.png]{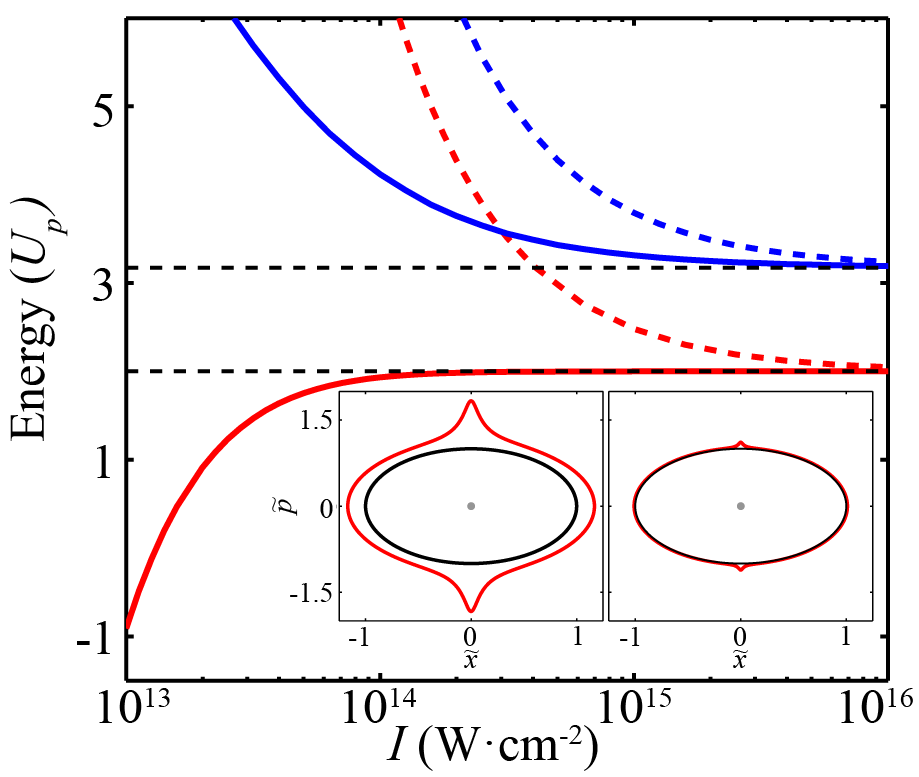}
 \caption{ \label{fig:maxReturn} Maximum return energy (solid lines) and maximum return kinetic energy (dashed lines) accessible to the dynamics (blue) and on the periodic orbit ${\cal O}$ (red). Insets:  Periodic orbit $\mathcal{O}$ (red curves) and the SFA counterpart (black curves).  The left panel corresponds to $I=10^{14}$ (left) and the right panel to $I=10^{15} \ {\rm W}\cdot{\rm cm}^{-2}$.  The {\it \~{x}} and {\it \~{p}} axes are scaled in units of $E_0 / \omega^2$ and $E_0 / \omega$, respectively.}
\end{figure}

First, in order to assess the importance of the Coulomb interaction in the recollision process, we examine in Fig.~\ref{fig:maxReturn}, the value of the maximum return energy (in units of $U_p$) accessible to the dynamics as intensity is varied (see also Ref.~\cite{Band05}). The solid curves correspond to the maximum return energy ${\cal E}_{\rm max}$ (defined as kinetic energy plus Coulomb potential), while the dashed lines correspond to the maximum return kinetic energy ${\cal K}_{\rm max}$.  The blue curves show the maximum return energies accessible to the dynamics while the red curves correspond to the return energies for specific initial conditions on the periodic orbit $\mathcal{O}$ discussed below. The maximum return energies ${\cal E}_{\rm max}$ (solid blue curve) are always larger than $3.17 ~U_p$ and converge to that number as intensity is increased.  This convergence to $3.17 ~U_p$ is expected, because with increasing intensity the laser field dominates over the Coulomb interaction and the SFA becomes valid. To measure the influence of the Coulomb interaction, the maximum return kinetic energy ${\cal K}_{\rm max}$ is plotted in Fig.~\ref{fig:maxReturn} in dashed lines. The two blue curves depart significantly from each other as intensity is decreased below $10^{15}\ {\rm W}\cdot{\rm cm}^{-2}$. The convergence is faster for the maximum return energy ${\cal E}_{\rm max}$ than it is for the maximum return kinetic energy ${\cal K}_{\rm max}$. This sounds paradoxical since the convergence of ${\cal E}_{\rm max}$ (which contains the Coulomb potential) is faster than ${\cal K}_{\rm max}$, towards a result which does not take into account the Coulomb interaction (e.g, by comparing the blue and dashed blue in Fig.~\ref{fig:maxReturn}). In any case, it is clear that the SFA is not valid below intensities on the order of $10^{15}\ {\rm W}\cdot{\rm cm}^{-2}$.
 
To resolve this paradox, we consider the classical dynamics of a recolliding electron in a strong linearly polarized laser field in the dipole approximation (and in atomic units)
\begin{equation} \label{eq:Hamiltonian_1e1d}
	{\mathcal{H}}\left({\bf x},{\bf p},t\right) = \frac{{\bf p}^2}{2} +V({\bf x})  + {\bf E}(t)\cdot {\bf x}, 
\end{equation}
where ${\bf E}(t)$ is the electric field (of intensity $I$), $\bf x$ the position of the electron and $\bf p$ its kinetic momentum. The shape of the ionic potential does not matter for the argumentation, provided that at long distance it vanishes sufficiently fast (e.g. $-1/\Vert {\bf x}\Vert$). We consider an electron leaving the core region at time $t=t_i$ and returning at $t=t_r$, i.e., ${\bf x}(t_r)={\bf 0}$. Integrating the dynamical equation for the kinetic momentum between ionization and recollision times, we have
$$
{\bf p}(t_r)-{\bf p}(t_i)=\Delta {\bf p}_{\rm E}+\Delta {\bf p}_{\rm C},
$$ 
where $\Delta {\bf p}_{\rm E}$ is the contribution in the absence of Coulomb potential
$$
\Delta {\bf p}_{\rm E}=-\int_{t_i}^{t_r} {\bf E}(t)dt,
$$
and $\Delta {\bf p}_{\rm C}$ is the change in kinetic momentum generated by the Coulomb potential
\begin{equation}
\label{eqn:intC}
\Delta {\bf p}_{\rm C}=-\int_{t_i}^{t_r} \nabla V({\bf x}(t))dt.
\end{equation}
The dominant contribution in the integral~(\ref{eqn:intC}) is right before the recollision time $t_r$. We expand ${\bf x}(t)$ around $t=t_r$ which leads to ${\bf x}(t)\approx {\bf p}(t_r)(t-t_r)$ at the recollision time. Inserting this approximation in the integral gives 
\begin{equation}
\label{eqn:cancel1}
{\bf p}(t_r)\cdot \Delta {\bf p}_{\rm C}\approx -V({\bf 0})+V({\bf p}(t_r)(t_i-t_r)).
\end{equation}
Since the potential vanishes far away from the ionic core and $\Vert {\bf p}(t_r)\Vert$ is large, the second term in the right hand side is negligible, of order $\omega^2/(\pi E_0)$, in comparison with $I_p$. 
In the SFA, $\Delta {\bf p}_{\rm C}$ is neglected. Here we go one order further and assume that $\Delta {\bf p}_{\rm C}$ is negligible in comparison with ${\bf p}(t_r)$. Looking at the kinetic energy and using Eq.~(\ref{eqn:cancel1}), we arrive at
\begin{equation}
\label{eqn:cancel2}
\frac{{\bf p}(t_r)^2}{2}+V({\bf 0})\approx \frac{{\bf p}_{\rm SFA}^2}{2},
\end{equation}
where ${\bf p}_{\rm SFA}={\bf p}(t_i)+\Delta {\bf p}_{\rm E}$ is the kinetic momentum in the SFA. The hypothesis involved in the derivation above translates into some admissible range of parameters where Eq.~(\ref{eqn:cancel2}) is expected to be valid:
$$
\frac{\omega^2}{\pi\sqrt{I} }\ll I_p\ll \frac{I}{\omega^2},
$$  
which is a regime of small Keldysh parameters~\cite{Ivan05}. 
Equation~(\ref{eqn:cancel2}) states that the maximum return energy for the recolliding electron is equal to the maximum return {\em kinetic} energy in the SFA.  In the intensity range from $10^{15}$ to $10^{16} \ {\rm W}\cdot{\rm cm}^{-2}$, it validates the results obtained in the SFA which make use of the maximum return kinetic energy and ignore the Coulomb potential. We notice that the tunneling argument is not needed to obtain $\kappa_0 U_p$ as the maximum return energy and the Coulomb potential has been fully taken into account. 

In the intermediate range of intensities, from $10^{14}$ to $10^{15} \ {\rm W}\cdot{\rm cm}^{-2}$, there are some deviations from the SFA value. An approximate value of $\kappa(I,\omega)$, the maximum return energy in units of $U_p$, is obtained by fitting the curve in Fig.~\ref{fig:maxReturn}:
$$
\kappa(I,\omega)\approx \kappa_0+\kappa_1\frac{\omega^2}{I},
$$
with some constant $\kappa_1$. This quantifies approximately the first-order effect of the Coulomb interaction on the maximum return energy at recollision as a function of the laser parameters. However, this correction is only sizable for low intensities (below $10^{14} \ {\rm W}\cdot{\rm cm}^{-2}$) and when the ponderomotive energy is also low, its influence cannot be easily identified on HHG cut-offs.
In this intermediate range of intensities, the three-step scenario is invalid and a correct recollision scenario needs to fully incorporate the effect of the ionic core.  We now turn to the dynamics of electrons driven by both fields. 

The fundamental question to be addressed is~: By what mechanism do almost-ionized electrons return to the core in a combined Coulomb and laser field? Each recolliding trajectory looks very different when visualized, and they seem to have nothing in common other than shuttling between the core and the far-field regions. However, we find that they do so by tracking a specific periodic orbit which has the same period as the laser field and represents the prototype of a recollision (see Fig.~\ref{fig:maxReturn}): A trajectory started close to this orbit (which we call $\mathcal{O}$) experiences large excursions from the core (beyond the quiver radius $E_0/\omega^2$) and returns to the core periodically, twice per laser period (with a momentum of order $E_0/\omega$ for large intensities, i.e., a maximum return kinetic energy of $2~U_p$). The maximum return energy is depicted by a continuous red curve on Fig.~\ref{fig:maxReturn} and it gives a natural explanation for the first cut-off in Fig.~\ref{fig:HHG} since the initial wave packet is initiated very close to $\mathcal{O}$. So at least for a short time, we see the imprint of ${\cal O}$ on the wave packet dynamics. For longer times, the wave packet explores a wider region in phase space, and therefore the maximum return energy prevails (blue curve in Fig.~\ref{fig:maxReturn}).  

\begin{figure}
 \centering
 \includegraphics[width = \linewidth, type=png, read=.png, ext=.png]{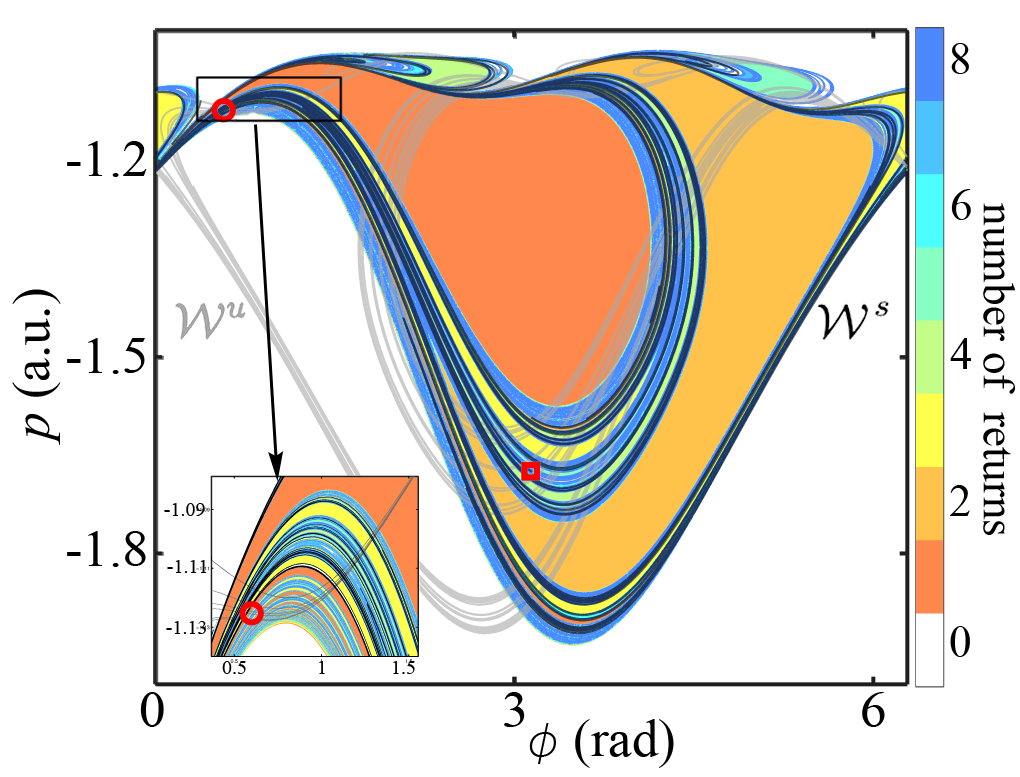}		 
 \caption{ \label{fig:unstableManifold} Stable (${\cal W}^s$, in black) and unstable ($\mathcal{W}^u$, in gray) manifolds of the periodic orbit $\mathcal{O}$ visualized on the Poincar\'{e} section $x=0$.  The intensity is $I=10^{14} \ {\rm W}\cdot{\rm cm}^{-2}$. The red square marker corresponds to the location of the periodic orbit $\mathcal{O}$ on the section.  The colored areas correspond to regions in phase space (on the section) leading to recollision and the color scale denotes the number of returns.  The trajectories initiated in the white region ionize without returning to the core, or remain bound by the Coulomb potential indefinitely.  The bounding box on the top left of the figure is the region shown in the inset.  The red circular marker denotes the initial condition which results in an electron returning with maximum energy.}
\end{figure}

Remarkably, linear stability analysis~\cite{chaosbook} shows that ${\cal O}$ is only weakly unstable  which means that trajectories can follow this periodic orbit long enough to be influenced by it and its stable and unstable manifolds, which channel ionizations and returns to the core. Electrons move away from $\mathcal{O}$ following its unstable manifold, $\mathcal{W}^u$, and return to this orbit following its stable manifold ${\cal W}^s$~\cite{chaosbook}. The two manifolds are linked by the symmetry $(x,p,\phi)\mapsto (-x,p,2\pi-\phi)$ when time reverses direction.  Both manifolds ${\cal W}^u$ and $\mathcal{W}^s$ are visualized in Fig.~\ref{fig:unstableManifold} in the Poincar\'e section $x=0$ in the plane $(\phi,p)$ in the lower half plane $p<0$. The chosen Poincar\'e section is the natural one for recollisions since, by definition, they occur at the core, $x=0$. We note that there are two types of recollisions: the ones which recollide with a positive momentum (from left to right) and the other ones with a negative momentum (from right to left). The upper half plane (left to right recollisions) is related to the lower half by the symmetry $(x,p,\phi)\mapsto (x,-p,\pi-\phi)$ upon time-reversal.  
In Fig.~\ref{fig:unstableManifold} we show the set of initial conditions on the surface $x=0$ (and lower half plane $p<0$) leading to subsequent recollisions. The color scale corresponds to the number of recollisions (regardless of the direction, right to left or left to right) undergone by the trajectory started with the given initial conditions. Remarkably, all these initial conditions are located around the stable manifold ${\cal W}^s$ and the higher the number of recollisions, the closer the initial conditions are to this manifold. This is a firm indication that $\cal O$ organizes the recollision dynamics through its stable manifold. The maximum return energy can also be found from an examination of the manifold: The largest momentum available to an electron moving along $\mathcal{W}^u$ before leaving the core corresponds to the maximum return energy, e.g., $4.31 ~U_p$ at $10^{14} \ {\rm W}\cdot{\rm cm}^{-2}$ and $3.32 ~U_p$ at $10^{15} \ {\rm W}\cdot{\rm cm}^{-2}$. These energies fall on the solid blue curve of Fig.~\ref{fig:maxReturn} which gives the maximum return energy allowed by the dynamics. The SFA relies on tunneling to release the photoelectron near the core (where the Coulomb field has dropped off) with zero momentum~\cite{Cork93,Band05}. With those initial conditions -- and under the influence of the laser field alone -- the SFA finds a single special trajectory which gives the well-known maximal return energy of $3.17~{ U_p}$. When the recollision is not broken up into steps (one with the Coulomb field and another without as in the SFA), but rather regarded as one continuous process with the Coulomb field always on, the notion of zero-initial momentum after tunneling loses its significance.  
More importantly, the maximum amount of energy an electron can bring back to the core upon recollision depends on the duration of the pulse. The analysis of the dynamics shows that the most energetic recollisions do not happen within the next half laser cycle after preionization but take much longer and, the longer the delay to return, the more energy they are likely to bring back.  
For instance, at the intensity of $10^{14} \ {\rm W}\cdot{\rm cm}^{-2}$, the maximum return energy is $3.86~U_p$ for recollisions taking less than one laser cycle, and $4.24~U_p$ when allowing up to $25$ laser cycles to return.
The area highlighted by the bounding box in Fig.~\ref{fig:unstableManifold} (and also shown in an expanded view in the inset) shows the initial condition (circular red marker) which ultimately returns with maximum energy for a 10 laser cycle pulse.
Overall, the highest return energy for recollision corresponds to the limit of a pulse with infinite length -- an impractical scenario.
Nevertheless, the differences between these maxima are not significant enough to be observed in HHG spectra:
If the ponderomotive energy is low, this difference corresponds to a very small number of harmonics in the spectrum (at $I=10^{14} \ {\rm W}\cdot{\rm cm}^{-2}$ the value of $U_p$ is approximately 4 harmonics).

In summary, we have built an internally consistent recollision picture by including the Coulomb interaction in the traditional recollision picture. We find that recollisions are driven by a specific periodic orbit born out of a resonance with the field. Its stable/unstable manifolds guide the trajectories away and back to the core and, with a remarkable cancellation nullifying the influence of the ionic core potential, provide a purely classical interpretation of the HHG spectra cut-offs. In the current scenario with the linearly polarized field, the specific role of the ion's Coulomb field has been obscured by the seeming success of the SFA to date, even though in the intermediate range of intensities, it is based on wrong assumptions. We anticipate that the mechanism shown in this work -- namely recollisions being driven and regulated by recolliding periodic orbits~\cite{Kamo13} and their associated manifolds -- extends to all polarizations or wave forms, which opens up a promising avenue to extend the harmonic cut-offs beyond their single-color limit~\cite{Haes13}.

A.K.\ acknowledges financial support from the Chateaubriand fellowship program of the Embassy of France in the United States. 
A.K.\ and T.U.\ acknowledge funding from the NSF. F.M.\ acknowledges financial support from the Centre de Recherches Math\'ematiques. 
The research leading to these results has received funding from the People Programme (Marie Curie Actions) of the European Union's Seventh Framework Programme FP7/2007-2013/ under REA grant agreement 294974. 


\begin{thebibliography}{25}
\expandafter\ifx\csname natexlab\endcsname\relax\def\natexlab#1{#1}\fi
\expandafter\ifx\csname bibnamefont\endcsname\relax
  \def\bibnamefont#1{#1}\fi
\expandafter\ifx\csname bibfnamefont\endcsname\relax
  \def\bibfnamefont#1{#1}\fi
\expandafter\ifx\csname citenamefont\endcsname\relax
  \def\citenamefont#1{#1}\fi
\expandafter\ifx\csname url\endcsname\relax
  \def\url#1{\texttt{#1}}\fi
\expandafter\ifx\csname urlprefix\endcsname\relax\def\urlprefix{URL }\fi
\providecommand{\bibinfo}[2]{#2}
\providecommand{\eprint}[2][]{\url{#2}}

\bibitem[{\citenamefont{Corkum and Krausz}(2007)}]{Cork07}
\bibinfo{author}{\bibfnamefont{P.~B.} \bibnamefont{Corkum}} \bibnamefont{and}
  \bibinfo{author}{\bibfnamefont{F.}~\bibnamefont{Krausz}},
  \bibinfo{journal}{Nature Physics} \textbf{\bibinfo{volume}{3}},
  \bibinfo{pages}{381 } (\bibinfo{year}{2007}), ISSN \bibinfo{issn}{17452473}.

\bibitem[{\citenamefont{Krausz and Ivanov}(2009)}]{Krau09}
\bibinfo{author}{\bibfnamefont{F.}~\bibnamefont{Krausz}} \bibnamefont{and}
  \bibinfo{author}{\bibfnamefont{M.}~\bibnamefont{Ivanov}},
  \bibinfo{journal}{Rev.~Mod.~Phys.} \textbf{\bibinfo{volume}{81}},
  \bibinfo{pages}{163} (\bibinfo{year}{2009}).

\bibitem[{\citenamefont{Itatani et~al.}(2004)\citenamefont{Itatani, Levesque,
  Zeidler, Niikura, Pepin, Kieffer, Corkum, and Villeneuve}}]{Itat04}
\bibinfo{author}{\bibfnamefont{J.}~\bibnamefont{Itatani}},
  \bibinfo{author}{\bibfnamefont{J.}~\bibnamefont{Levesque}},
  \bibinfo{author}{\bibfnamefont{D.}~\bibnamefont{Zeidler}},
  \bibinfo{author}{\bibfnamefont{H.}~\bibnamefont{Niikura}},
  \bibinfo{author}{\bibfnamefont{H.}~\bibnamefont{Pepin}},
  \bibinfo{author}{\bibfnamefont{J.~C.} \bibnamefont{Kieffer}},
  \bibinfo{author}{\bibfnamefont{P.~B.} \bibnamefont{Corkum}},
  \bibnamefont{and} \bibinfo{author}{\bibfnamefont{D.~M.}
  \bibnamefont{Villeneuve}}, \bibinfo{journal}{Nature}
  \textbf{\bibinfo{volume}{432}}, \bibinfo{pages}{867} (\bibinfo{year}{2004}).

\bibitem[{\citenamefont{Corkum}(1993)}]{Cork93}
\bibinfo{author}{\bibfnamefont{P.~B.} \bibnamefont{Corkum}},
  \bibinfo{journal}{Phys.~Rev.~Lett.} \textbf{\bibinfo{volume}{71}},
  \bibinfo{pages}{1994} (\bibinfo{year}{1993}).

\bibitem[{\citenamefont{Schafer et~al.}(1993)\citenamefont{Schafer, Yang,
  DiMauro, and Kulander}}]{Scha93}
\bibinfo{author}{\bibfnamefont{K.~J.} \bibnamefont{Schafer}},
  \bibinfo{author}{\bibfnamefont{B.}~\bibnamefont{Yang}},
  \bibinfo{author}{\bibfnamefont{L.~F.} \bibnamefont{DiMauro}},
  \bibnamefont{and} \bibinfo{author}{\bibfnamefont{K.~C.}
  \bibnamefont{Kulander}}, \bibinfo{journal}{Phys.~Rev.~Lett.}
  \textbf{\bibinfo{volume}{70}}, \bibinfo{pages}{1599} (\bibinfo{year}{1993}).

\bibitem[{\citenamefont{Becker et~al.}(2012)\citenamefont{Becker, Liu, Ho, and
  Eberly}}]{Beck12}
\bibinfo{author}{\bibfnamefont{W.}~\bibnamefont{Becker}},
  \bibinfo{author}{\bibfnamefont{X.}~\bibnamefont{Liu}},
  \bibinfo{author}{\bibfnamefont{P.~J.} \bibnamefont{Ho}}, \bibnamefont{and}
  \bibinfo{author}{\bibfnamefont{J.~H.} \bibnamefont{Eberly}},
  \bibinfo{journal}{Rev.~Mod.~Phys.} \textbf{\bibinfo{volume}{84}},
  \bibinfo{pages}{1011} (\bibinfo{year}{2012}).

\bibitem[{\citenamefont{Lewenstein et~al.}(1994)\citenamefont{Lewenstein,
  Balcou, Ivanov, L'Huillier, and Corkum}}]{Lewe94}
\bibinfo{author}{\bibfnamefont{M.}~\bibnamefont{Lewenstein}},
  \bibinfo{author}{\bibfnamefont{P.}~\bibnamefont{Balcou}},
  \bibinfo{author}{\bibfnamefont{M.~Y.} \bibnamefont{Ivanov}},
  \bibinfo{author}{\bibfnamefont{A.}~\bibnamefont{L'Huillier}},
  \bibnamefont{and} \bibinfo{author}{\bibfnamefont{P.~B.}
  \bibnamefont{Corkum}}, \bibinfo{journal}{Phys.~Rev.~A}
  \textbf{\bibinfo{volume}{49}}, \bibinfo{pages}{2117} (\bibinfo{year}{1994}).

\bibitem[{\citenamefont{L'Huillier et~al.}(1993)\citenamefont{L'Huillier,
  Lewenstein, Sali\`eres, Balcou, Ivanov, Larsson, and Wahlstr\"om}}]{Lhui93b}
\bibinfo{author}{\bibfnamefont{A.}~\bibnamefont{L'Huillier}},
  \bibinfo{author}{\bibfnamefont{M.}~\bibnamefont{Lewenstein}},
  \bibinfo{author}{\bibfnamefont{P.}~\bibnamefont{Sali\`eres}},
  \bibinfo{author}{\bibfnamefont{P.}~\bibnamefont{Balcou}},
  \bibinfo{author}{\bibfnamefont{M.~Y.} \bibnamefont{Ivanov}},
  \bibinfo{author}{\bibfnamefont{J.}~\bibnamefont{Larsson}}, \bibnamefont{and}
  \bibinfo{author}{\bibfnamefont{C.~G.} \bibnamefont{Wahlstr\"om}},
  \bibinfo{journal}{Phys. Rev. A} \textbf{\bibinfo{volume}{48}},
  \bibinfo{pages}{R3433} (\bibinfo{year}{1993}).

\bibitem[{\citenamefont{Krause et~al.}(1992)\citenamefont{Krause, Schafer, and
  Kulander}}]{Krau92}
\bibinfo{author}{\bibfnamefont{J.~L.} \bibnamefont{Krause}},
  \bibinfo{author}{\bibfnamefont{K.~J.} \bibnamefont{Schafer}},
  \bibnamefont{and} \bibinfo{author}{\bibfnamefont{K.~C.}
  \bibnamefont{Kulander}}, \bibinfo{journal}{Phys. Rev. Lett.}
  \textbf{\bibinfo{volume}{68}}, \bibinfo{pages}{3535} (\bibinfo{year}{1992}).

\bibitem[{\citenamefont{Protopapas et~al.}(1996)\citenamefont{Protopapas,
  Lappas, Keitel, and Knight}}]{Prot96}
\bibinfo{author}{\bibfnamefont{M.}~\bibnamefont{Protopapas}},
  \bibinfo{author}{\bibfnamefont{D.~G.} \bibnamefont{Lappas}},
  \bibinfo{author}{\bibfnamefont{C.~H.} \bibnamefont{Keitel}},
  \bibnamefont{and} \bibinfo{author}{\bibfnamefont{P.~L.}
  \bibnamefont{Knight}}, \bibinfo{journal}{Phys. Rev. A}
  \textbf{\bibinfo{volume}{53}}, \bibinfo{pages}{R2933} (\bibinfo{year}{1996}).

\bibitem[{\citenamefont{van~de Sand and Rost}(1999)}]{Sand99}
\bibinfo{author}{\bibfnamefont{G.}~\bibnamefont{van~de Sand}} \bibnamefont{and}
  \bibinfo{author}{\bibfnamefont{J.~M.} \bibnamefont{Rost}},
  \bibinfo{journal}{Phys.~Rev.~Lett.} \textbf{\bibinfo{volume}{83}},
  \bibinfo{pages}{524} (\bibinfo{year}{1999}).

\bibitem[{\citenamefont{L'Huillier and Balcou}(1993)}]{Lhui93a}
\bibinfo{author}{\bibfnamefont{A.}~\bibnamefont{L'Huillier}} \bibnamefont{and}
  \bibinfo{author}{\bibfnamefont{P.}~\bibnamefont{Balcou}},
  \bibinfo{journal}{Phys. Rev. Lett.} \textbf{\bibinfo{volume}{70}},
  \bibinfo{pages}{774} (\bibinfo{year}{1993}).

\bibitem[{\citenamefont{Pfeiffer et~al.}(2012)\citenamefont{Pfeiffer, Cirelli,
  Smolarski, Dimitrovski, Abu-samha, Madsen, and Keller}}]{Pfei12}
\bibinfo{author}{\bibfnamefont{A.~N.} \bibnamefont{Pfeiffer}},
  \bibinfo{author}{\bibfnamefont{C.}~\bibnamefont{Cirelli}},
  \bibinfo{author}{\bibfnamefont{M.}~\bibnamefont{Smolarski}},
  \bibinfo{author}{\bibfnamefont{D.}~\bibnamefont{Dimitrovski}},
  \bibinfo{author}{\bibfnamefont{M.}~\bibnamefont{Abu-samha}},
  \bibinfo{author}{\bibfnamefont{L.~B.} \bibnamefont{Madsen}},
  \bibnamefont{and} \bibinfo{author}{\bibfnamefont{U.}~\bibnamefont{Keller}},
  \bibinfo{journal}{Nature Physics} \textbf{\bibinfo{volume}{8}},
  \bibinfo{pages}{76 } (\bibinfo{year}{2012}), ISSN \bibinfo{issn}{17452473}.

\bibitem[{\citenamefont{Shafir et~al.}(2012{\natexlab{a}})\citenamefont{Shafir,
  Soifer, Bruner, Dagan, Mairesse, Patchkovskii, Ivanov, Smirnova, and
  Dudovich}}]{Shaf12b}
\bibinfo{author}{\bibfnamefont{D.}~\bibnamefont{Shafir}},
  \bibinfo{author}{\bibfnamefont{H.}~\bibnamefont{Soifer}},
  \bibinfo{author}{\bibfnamefont{B.~D.} \bibnamefont{Bruner}},
  \bibinfo{author}{\bibfnamefont{M.}~\bibnamefont{Dagan}},
  \bibinfo{author}{\bibfnamefont{Y.}~\bibnamefont{Mairesse}},
  \bibinfo{author}{\bibfnamefont{S.}~\bibnamefont{Patchkovskii}},
  \bibinfo{author}{\bibfnamefont{M.~Y.} \bibnamefont{Ivanov}},
  \bibinfo{author}{\bibfnamefont{O.}~\bibnamefont{Smirnova}}, \bibnamefont{and}
  \bibinfo{author}{\bibfnamefont{N.}~\bibnamefont{Dudovich}},
  \bibinfo{journal}{Nature} \textbf{\bibinfo{volume}{485}}, \bibinfo{pages}{343
  } (\bibinfo{year}{2012}{\natexlab{a}}), ISSN \bibinfo{issn}{1476-4687}.

\bibitem[{\citenamefont{Ivanov et~al.}(2005)\citenamefont{Ivanov, Spanner, and
  Smirnova}}]{Ivan05}
\bibinfo{author}{\bibfnamefont{M.~Y.} \bibnamefont{Ivanov}},
  \bibinfo{author}{\bibfnamefont{M.}~\bibnamefont{Spanner}}, \bibnamefont{and}
  \bibinfo{author}{\bibfnamefont{O.}~\bibnamefont{Smirnova}},
  \bibinfo{journal}{J.~Mod.~Opt.} \textbf{\bibinfo{volume}{52}},
  \bibinfo{pages}{165} (\bibinfo{year}{2005}).

\bibitem[{\citenamefont{Brabec et~al.}(1996)\citenamefont{Brabec, Ivanov, and
  Corkum}}]{Barb96}
\bibinfo{author}{\bibfnamefont{T.}~\bibnamefont{Brabec}},
  \bibinfo{author}{\bibfnamefont{M.~Y.} \bibnamefont{Ivanov}},
  \bibnamefont{and} \bibinfo{author}{\bibfnamefont{P.~B.}
  \bibnamefont{Corkum}}, \bibinfo{journal}{Phys.~Rev.~A}
  \textbf{\bibinfo{volume}{54}}, \bibinfo{pages}{R2551} (\bibinfo{year}{1996}).

\bibitem[{\citenamefont{Shafir et~al.}(2013)\citenamefont{Shafir, Soifer,
  Vozzi, Johnson, Hartung, Dube, Villeneuve, Corkum, Dudovich, and
  Staudte}}]{Shaf13}
\bibinfo{author}{\bibfnamefont{D.}~\bibnamefont{Shafir}},
  \bibinfo{author}{\bibfnamefont{H.}~\bibnamefont{Soifer}},
  \bibinfo{author}{\bibfnamefont{C.}~\bibnamefont{Vozzi}},
  \bibinfo{author}{\bibfnamefont{A.~S.} \bibnamefont{Johnson}},
  \bibinfo{author}{\bibfnamefont{A.}~\bibnamefont{Hartung}},
  \bibinfo{author}{\bibfnamefont{Z.}~\bibnamefont{Dube}},
  \bibinfo{author}{\bibfnamefont{D.~M.} \bibnamefont{Villeneuve}},
  \bibinfo{author}{\bibfnamefont{P.~B.} \bibnamefont{Corkum}},
  \bibinfo{author}{\bibfnamefont{N.}~\bibnamefont{Dudovich}}, \bibnamefont{and}
  \bibinfo{author}{\bibfnamefont{A.}~\bibnamefont{Staudte}},
  \bibinfo{journal}{Phys. Rev. Lett.} \textbf{\bibinfo{volume}{111}},
  \bibinfo{pages}{023005} (\bibinfo{year}{2013}).

\bibitem[{\citenamefont{Li et~al.}(2013)\citenamefont{Li, Liu, Liu, Ning, Fu,
  Liu, Deng, Wu, Peng, and Gong}}]{Li13}
\bibinfo{author}{\bibfnamefont{M.}~\bibnamefont{Li}},
  \bibinfo{author}{\bibfnamefont{Y.}~\bibnamefont{Liu}},
  \bibinfo{author}{\bibfnamefont{H.}~\bibnamefont{Liu}},
  \bibinfo{author}{\bibfnamefont{Q.}~\bibnamefont{Ning}},
  \bibinfo{author}{\bibfnamefont{L.}~\bibnamefont{Fu}},
  \bibinfo{author}{\bibfnamefont{J.}~\bibnamefont{Liu}},
  \bibinfo{author}{\bibfnamefont{Y.}~\bibnamefont{Deng}},
  \bibinfo{author}{\bibfnamefont{C.}~\bibnamefont{Wu}},
  \bibinfo{author}{\bibfnamefont{L.}~\bibnamefont{Peng}}, \bibnamefont{and}
  \bibinfo{author}{\bibfnamefont{Q.}~\bibnamefont{Gong}},
  \bibinfo{journal}{Phys. Rev. Lett.} \textbf{\bibinfo{volume}{111}},
  \bibinfo{pages}{023006} (\bibinfo{year}{2013}).

\bibitem[{\citenamefont{Shafir et~al.}(2012{\natexlab{b}})\citenamefont{Shafir,
  Fabre, Higuet, Soifer, Dagan, Descamps, M\'evel, Petit, W\"orner, Pons
  et~al.}}]{Shaf12a}
\bibinfo{author}{\bibfnamefont{D.}~\bibnamefont{Shafir}},
  \bibinfo{author}{\bibfnamefont{B.}~\bibnamefont{Fabre}},
  \bibinfo{author}{\bibfnamefont{J.}~\bibnamefont{Higuet}},
  \bibinfo{author}{\bibfnamefont{H.}~\bibnamefont{Soifer}},
  \bibinfo{author}{\bibfnamefont{M.}~\bibnamefont{Dagan}},
  \bibinfo{author}{\bibfnamefont{D.}~\bibnamefont{Descamps}},
  \bibinfo{author}{\bibfnamefont{E.}~\bibnamefont{M\'evel}},
  \bibinfo{author}{\bibfnamefont{S.}~\bibnamefont{Petit}},
  \bibinfo{author}{\bibfnamefont{H.~J.} \bibnamefont{W\"orner}},
  \bibinfo{author}{\bibfnamefont{B.}~\bibnamefont{Pons}}, \bibnamefont{et~al.},
  \bibinfo{journal}{Phys. Rev. Lett.} \textbf{\bibinfo{volume}{108}},
  \bibinfo{pages}{203001} (\bibinfo{year}{2012}{\natexlab{b}}).

\bibitem[{\citenamefont{Javanainen et~al.}(1988)\citenamefont{Javanainen,
  Eberly, and Su}}]{Java88}
\bibinfo{author}{\bibfnamefont{J.}~\bibnamefont{Javanainen}},
  \bibinfo{author}{\bibfnamefont{J.~H.} \bibnamefont{Eberly}},
  \bibnamefont{and} \bibinfo{author}{\bibfnamefont{Q.}~\bibnamefont{Su}},
  \bibinfo{journal}{Phys.~Rev.~A} \textbf{\bibinfo{volume}{38}},
  \bibinfo{pages}{3430} (\bibinfo{year}{1988}).

\bibitem[{\citenamefont{Bleda et~al.}(2013)\citenamefont{Bleda, Yavuz, Altun,
  and Topcu}}]{Bled13}
\bibinfo{author}{\bibfnamefont{E.}~\bibnamefont{Bleda}},
  \bibinfo{author}{\bibfnamefont{I.}~\bibnamefont{Yavuz}},
  \bibinfo{author}{\bibfnamefont{Z.}~\bibnamefont{Altun}}, \bibnamefont{and}
  \bibinfo{author}{\bibfnamefont{T.}~\bibnamefont{Topcu}}
  (\bibinfo{year}{2013}), \bibinfo{note}{arxiv:1308.6305v1 [physics.atom-ph]}.

\bibitem[{\citenamefont{Bandrauk et~al.}(2005)\citenamefont{Bandrauk,
  Chelkowski, and Goudreau}}]{Band05}
\bibinfo{author}{\bibfnamefont{A.~D.} \bibnamefont{Bandrauk}},
  \bibinfo{author}{\bibfnamefont{S.}~\bibnamefont{Chelkowski}},
  \bibnamefont{and} \bibinfo{author}{\bibfnamefont{S.}~\bibnamefont{Goudreau}},
  \bibinfo{journal}{J.~Mod.~Opt.} \textbf{\bibinfo{volume}{52}},
  \bibinfo{pages}{411 } (\bibinfo{year}{2005}).

\bibitem[{\citenamefont{Cvitanovi\'{c}
  et~al.}(2008)\citenamefont{Cvitanovi\'{c}, Artuso, Mainieri, Tanner, and
  Vattay}}]{chaosbook}
\bibinfo{author}{\bibfnamefont{P.}~\bibnamefont{Cvitanovi\'{c}}},
  \bibinfo{author}{\bibfnamefont{R.}~\bibnamefont{Artuso}},
  \bibinfo{author}{\bibfnamefont{R.}~\bibnamefont{Mainieri}},
  \bibinfo{author}{\bibfnamefont{G.}~\bibnamefont{Tanner}}, \bibnamefont{and}
  \bibinfo{author}{\bibfnamefont{G.}~\bibnamefont{Vattay}},
  \emph{\bibinfo{title}{Chaos: Classical and Quantum}}
  (\bibinfo{publisher}{Niels Bohr Institute}, \bibinfo{address}{Copenhagen},
  \bibinfo{year}{2008}), \bibinfo{note}{\url{http://ChaosBook.org}}.

\bibitem[{\citenamefont{Kamor et~al.}(2013)\citenamefont{Kamor, Mauger,
  Chandre, and Uzer}}]{Kamo13}
\bibinfo{author}{\bibfnamefont{A.}~\bibnamefont{Kamor}},
  \bibinfo{author}{\bibfnamefont{F.}~\bibnamefont{Mauger}},
  \bibinfo{author}{\bibfnamefont{C.}~\bibnamefont{Chandre}}, \bibnamefont{and}
  \bibinfo{author}{\bibfnamefont{T.}~\bibnamefont{Uzer}},
  \bibinfo{journal}{Phys. Rev. Lett.} \textbf{\bibinfo{volume}{110}},
  \bibinfo{pages}{253002} (\bibinfo{year}{2013}).

\bibitem[{\citenamefont{Haessler et~al.}(2013)\citenamefont{Haessler,
  Balciunas, Fan, Witting, Squibb, Chipperfield, Zair, Andriukaitis, Pugzlys,
  Tisch et~al.}}]{Haes13}
\bibinfo{author}{\bibfnamefont{S.}~\bibnamefont{Haessler}},
  \bibinfo{author}{\bibfnamefont{T.}~\bibnamefont{Balciunas}},
  \bibinfo{author}{\bibfnamefont{G.}~\bibnamefont{Fan}},
  \bibinfo{author}{\bibfnamefont{T.}~\bibnamefont{Witting}},
  \bibinfo{author}{\bibfnamefont{R.}~\bibnamefont{Squibb}},
  \bibinfo{author}{\bibfnamefont{L.}~\bibnamefont{Chipperfield}},
  \bibinfo{author}{\bibfnamefont{A.}~\bibnamefont{Zair}},
  \bibinfo{author}{\bibfnamefont{G.}~\bibnamefont{Andriukaitis}},
  \bibinfo{author}{\bibfnamefont{A.}~\bibnamefont{Pugzlys}},
  \bibinfo{author}{\bibfnamefont{J.~W.~G.} \bibnamefont{Tisch}},
  \bibnamefont{et~al.} (\bibinfo{year}{2013}), \bibinfo{note}{arXiv:1308.5510
  [physics.atom-ph]}.

\end{thebibliography}

\end{document}